\documentclass{sigchi-ext}
\usepackage[T1]{fontenc}
\usepackage{textcomp}
\usepackage[scaled=.92]{helvet} 
\usepackage{color}
\usepackage{graphicx} 
\usepackage{balance}  
\usepackage{booktabs} 
\usepackage{ccicons}  
\usepackage{ragged2e} 

\usepackage{float}

\def\plaintitle{PlutoAR} \def\plainauthor{First Author, Second Author, Third Author,
  Fourth Author}
\def\emptyauthor{}
\def\plainkeywords{Augmented Reality; Interpreter, K-10 Education; User Interfaces}

\title{PlutoAR: An Inexpensive, Interactive And Portable Augmented Reality Based Interpreter For K-10 Curriculum}

\numberofauthors{4}
\author{
  \alignauthor{
    \textbf{Shourya Pratap Singh}\\
    \affaddr{International Institute of Information Technology} \\
    \affaddr{Bhubaneswar \\ Odisha, India} \\
    \email{b114040@iiit-bh.ac.in} } \alignauthor{
    \textbf{Ankit Kumar Panda}\\
     \affaddr{International Institute of Information Technology} \\
    \affaddr{Bhubaneswar \\ Odisha, India} \\
    \email{b114004@iiit-bh.ac.in} } \vfil \alignauthor{
    \textbf{Susobhit Panigrahi}\\
     \affaddr{International Institute of Information Technology} \\
    \affaddr{Bhubaneswar \\ Odisha, India} \\
    \email{b114050@iiit-bh.ac.in} }
    \alignauthor{
    \textbf{Ajaya Kumar Dash}\\
    \affaddr{International Institute of Information Technology} \\
    \affaddr{Bhubaneswar \\ Odisha, India} \\
    \email{ajaya@iiit-bh.ac.in} } \vfil 
    \alignauthor{
    \textbf{Debi Prosad Dogra}\\
    \affaddr{Indian Institute of Technology} \\
    \affaddr{Bhubaneswar \\ Odisha, India} \\
    \email{dpdogra@iitbbs.ac.in} } \vfil
     }

\definecolor{linkColor}{RGB}{6,125,233}
\hypersetup{
  pdftitle={\plaintitle},
  pdfauthor={\plainauthor},
  pdfauthor={\emptyauthor},
  pdfkeywords={\plainkeywords},
  bookmarksnumbered,
  pdfstartview={FitH},
  colorlinks,
  citecolor=black,
  filecolor=black,
  linkcolor=black,
  urlcolor=linkColor,
  breaklinks=true,
}

\begin{document}

\copyrightinfo{}

\maketitle

\RaggedRight{} 

\begin{abstract}
  The regular K-10 curriculums often do not get the necessary of affordable technology involving interactive ways of teaching the prescribed curriculum with effective analytical skill building. 
In this paper, we present ``PlutoAR'', a paper-based augmented reality interpreter which is scalable, affordable, portable and can be used as a platform for skill building for the kids.  PlutoAR manages to overcome the conventional albeit non-interactive ways of teaching by incorporating augmented reality (AR) through an interactive toolkit to provide students the best of both worlds. Students cut out paper ``tiles''  and place these tiles one by one on a larger paper surface called ``Launchpad'' and use the PlutoAR mobile application which runs on any Android device with a camera and uses augmented reality to output each step of the program like an interpreter. PlutoAR has inbuilt AR experiences like stories, maze solving using conditional loops, simple elementary mathematics and the intuition of gravity.\end{abstract}
\keywords{\plainkeywords}

\category{H.5.2}{Information interfaces and presentation (e.g.,
  HCI)}{User Interfaces}
\pagebreak 
  
\section{Introduction}

The curriculum and teaching methodology in schools separates programming from other classes, disconnected from students' interest as told by Mitchel Resnick, MIT \cite{Soloway:1993:WTS:163430.164061}. He also added, we need to rethink how we introduce programming to students in order to overcome the sad state of programming in schools.

\subsection{Computational Thinking:}

Computational thinking (CT) is slowly paving its way into becoming a crucial skill for the students. First coined by Wing \cite{Wing:2006:CT:1118178.1118215}, 
CT involves solving problems, designing systems, and understanding human behaviour, by drawing on the concepts fundamental to computer science. 

CT improves the analytical ability of children and is being viewed as the core of all science, technology, engineering and mathematics (STEM) disciplines suggested by Henderson et al. \cite{Henderson:2007:CT:1227504.1227378}.
Programming has a constructive and measurable effect on children's achievement, not only in areas in STEM, but also in language skills, creativity, and social emotional interaction \cite{Lee:2011:CTY:1929887.1929902}.

 \subsection{Intrinsic Motivation:}
 According to Ryan \& Deci's ``self determination theory'' \cite{m._l._2000}, intrinsic motivation means performing some task for the sake of the task itself and not to accomplish extrinsic rewards for it. Our visits to schools and interacting with the children showed intrinsic motivation in the maximum of children to learn with PlutoAR. The school visits have been elaborated ahead in the paper.
 \subsection{AR in Education:}
	Azuma \cite{doi:10.1162/pres.1997.6.4.355} defined Augmented Reality (AR) as a variation of Virtual Environment, that allows user to experience the seamless integration of virtual objects, superimposed upon or composited, with the real world. 
Hence, AR supplements reality, rather than completely replacing it. 

As advancement in wireless technology has paved the way for smaller and faster gadgets during the last three years, developers have found it easier to move AR from military and engineering applications to consumer uses \cite{5353455}. Also, the availability of mobile hardware supporting AR has become affordable and widespread. 

Johnson \cite{johnson_stone_smith_levine} stated that AR has strong potential to provide powerful contextual learning experiences.
Kirkpatrick \cite{kirkpatrick_1894} asserted that if visuals are shown to children, they can obtain a genuine knowledge of things more readily than they can be crammed with the verbal appearance of knowledge. Shelton and Hedley \cite{shelton_hedley_2004} suggested that AR not only creates visual images, but also has compelling possibilities for advancing spatial visualization. Spacial visualization can be useful in understanding concepts which cannot be possibly visualized and understood by simple visuals. The traditional approach for spacial visualization is to keep physical models, but they lack portability. With the advent of AR in mobile devices, AR systems have become completely portable. Chen \cite{Chen:2006:SCU:1128923.1128990} stated that because of portability, AR is more convenient to prepare as a supplement tool either in the classroom setting or at home.

Although AR is good at visualizing difficult concepts, visualization is not enough to enhance learning. In a research conducted by Kerawalla et al. \cite{Kerawalla:2006:XRE:1183174.1183176}, students taught using role-play approach learned more in comparison to students who watched animations in AR. Jablon and Wilkinson \cite{jablon2006using} stated that use of engagement strategies benefit children which  explains the reason behind these results. It is important for students to get involved in activities while learning. PlutoAR makes the best use of both worlds by making AR visualization an activity.

\subsection{Tangible Programming Kit:}
PlutoAR is a tangible programming kit. ``Tangible programming'', first coined by Suzuki and Kato \cite{Suzuki:1995:ISC:222020.222828} to describe their AlgoBlock programming environment for children. Tangible programming refers to the activity of arranging the graspable blocks  to build (as opposed to ``write'') computer programs \cite{mcnerney_2000}.

Current curriculum focuses too much on syntax. The environment provided to most introductory programming students forces them to get the syntax right before getting any feedback on their algorithm. Thus, the big picture of programming degenerates to an exercise in ``syntax satisfaction'' as put forward by Crews and Ziegler \cite{736854}. PlutoAR has no syntax, rather it has building blocks of program called ``tiles'' that let students see their program as a sequence of steps. This allows students to get actively involved in the process of design and creation, and learn more as suggested by Papert \cite{Papert:1980:MCC:1095592}.

PlutoAR has inbuilt AR experiences like stories such as taking off a rocket with facts about  the rocket to make children learn while they play. Maria Roussou says, ``play, in particular, can unite imagination and intellect in more than one way and help children discover things at their own pace and in their own way''. She also adds, ``play is a child's favourite activity, so the learning occurs more readily in an environment of fun, challenge, and variety'' \cite{Roussou:2004:LDL:973801.973818}. 

  \begin{marginfigure}[-26pc]
  \begin{minipage}{\marginparwidth}
    \centering
    \includegraphics[width=1.04\marginparwidth]{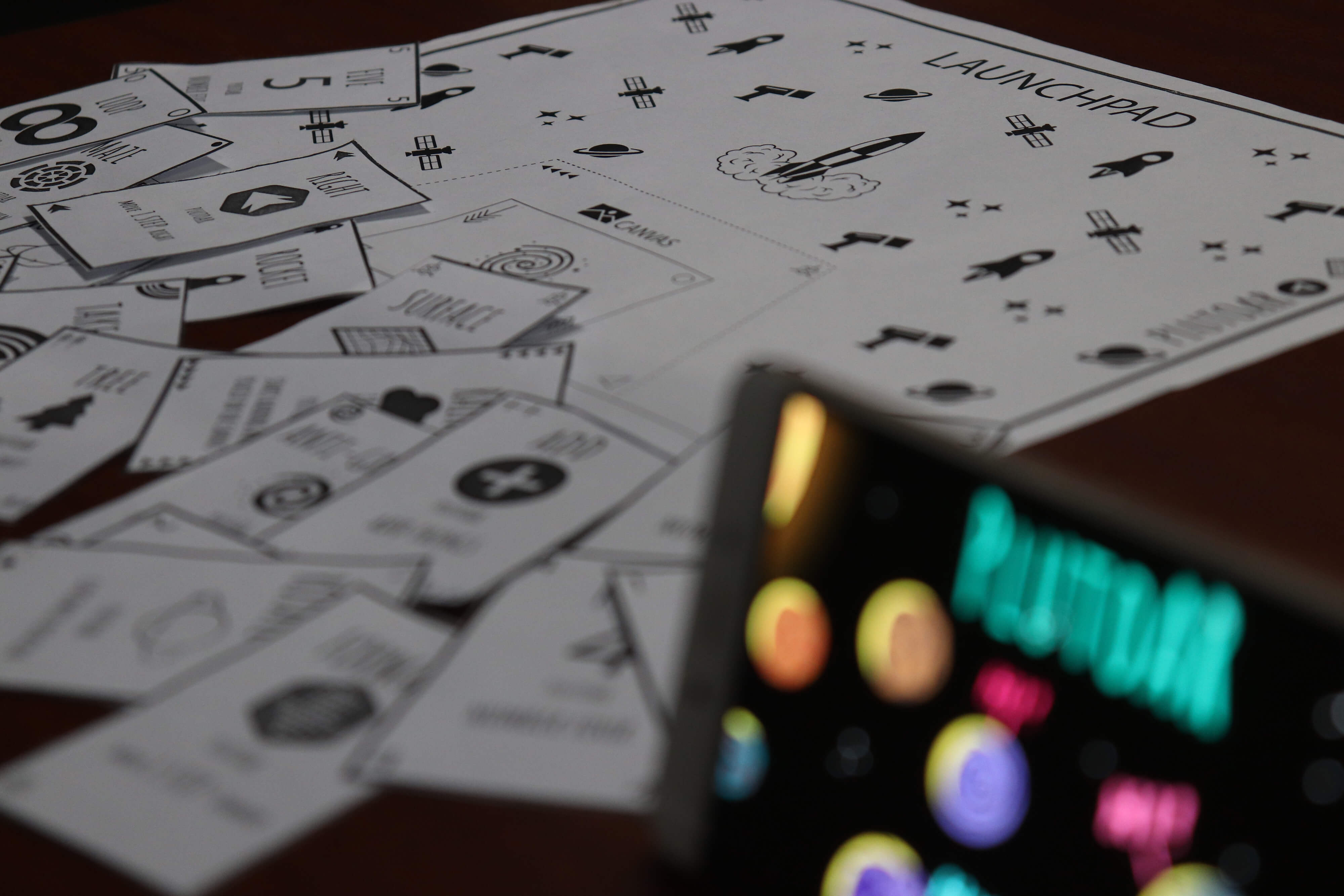}
    \caption{{The PlutoAR DreamKit consisting of Tiles, Launchpad, and PlutoAR Android Application.}}~\label{fig:marginfig2}
  \end{minipage}
\end{marginfigure}

  \begin{marginfigure}[-4pc]
  \begin{minipage}{\marginparwidth}
    \centering
    \includegraphics[width=1.04\marginparwidth]{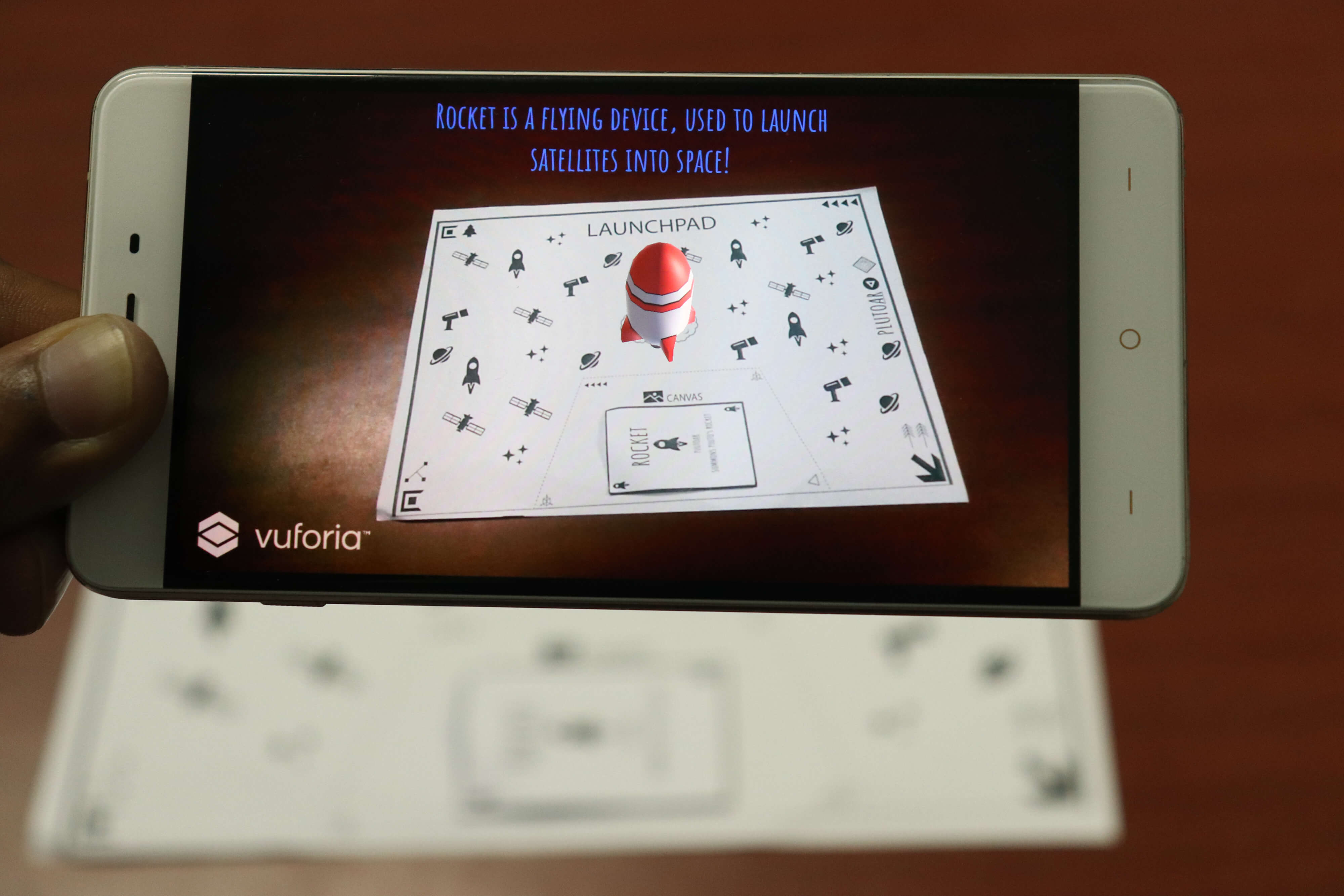}
    \caption{{Taking Off Rocket and learning facts about it in PlutoAR.}}~\label{fig:marginfig1}
  \end{minipage}
\end{marginfigure}

\section{Related Work}

A lot of research has gone into integrating Tangible Interface with AR to help children learn in much more immersive way. AR Scratch \cite{Radu:2009:ASC:1551788.1551831} is a popular software that adds AR with Scratch programming, enabling children to mix virtual stuff with reality. Tern \cite{Horn:2007:DTP:1226969.1227003}  is a tangible programming language used to teach computer programming to middle school students, by  joining wooden blocks as the program's building blocks. Daqri Elements4D   \cite{daqri} is another interactive AR solution that teaches basic chemistry by making use of cubes to show chemical reactions. Another toolkit named AlgoBlock \cite{Suzuki:1995:ISC:222020.222828} uses physical blocks that are joined together and later connected together to execute the program. CodeBits \cite{goyal2016code} helps to instill the idea of problem solving in kids by making use of paper bits put together in particular order to solve a maze.  
Most of today's AR Educational toolkits require heavy computing power and sophisticated sensors and accessories to achieve good results, which makes it costly, less portable, and non-mobile. PlutoAR, a computationally light-weight AR interpreter is mainly made to counter the above challenges by making learning inexpensive, immersive, portable and safe for children. It uses Android based mobile AR technology, helping kids to learn basics of programming, arithmetic and computational thinking. We strongly believe that technology in schools must be systematically supported for inculcating a teamwork effort amongst students.

\section{PlutoAR: Approach and Design}

PlutoAR comes as a ``DreamKit'' consisting of tiles, the launchpad and the Android mobile application. 

The ``tiles'' or programming blocks, are paper cut outs as shown in the Figure \ref{fig:marginfig2}. These make up the tangible aspect of PlutoAR. Children place these programming blocks according to the logic and imagination they have thought of for creating AR experiences. 

The launchpad is a large sheet of paper usually A4 in size which has two distinct regions outlined as ``canvas'' and ``space''. The ``canvas'' is the area where children place the ``tiles'' in a sequence and the interpreted result shows up as a 3D model when looked through PlutoAR's mobile application. The tiles and the launchpad are designed to be accessible and affordable. Thus, they are printed in black and white.
  \begin{marginfigure}[8pc]
  \begin{minipage}{\marginparwidth}
    \centering
    \includegraphics[width=1.04\marginparwidth]{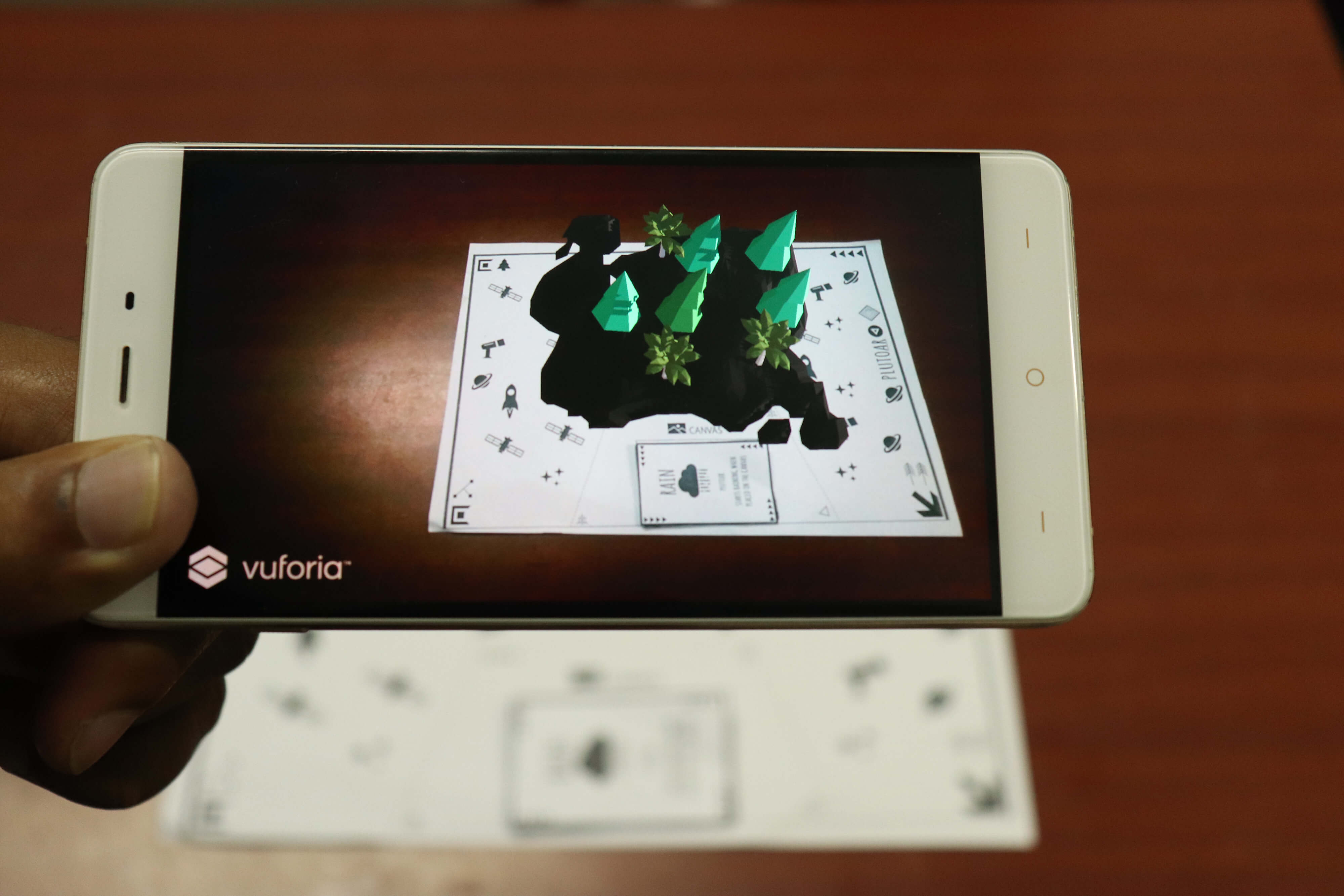}
    \caption{Using ``Rain'' tile to grow trees in AR.}~\label{fig:marginfig3}
  \end{minipage}
\end{marginfigure}

  \begin{marginfigure}[5pc]
  \begin{minipage}{\marginparwidth}
    \centering
    \includegraphics[width=1.04\marginparwidth]{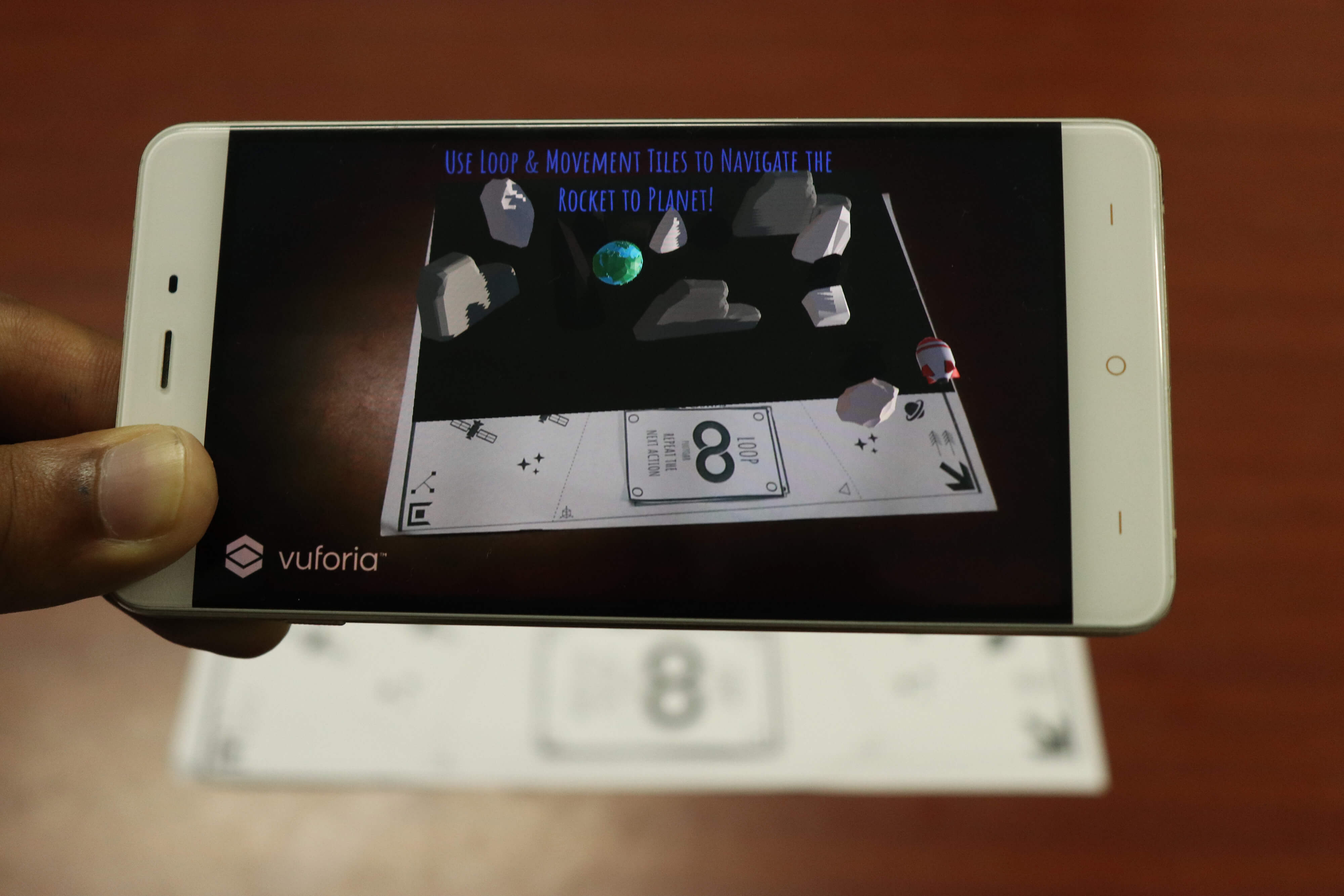}
    \caption{{Using ``Loop'' and Movement tiles to solve the maze.}}~\label{fig:marginfig4}
  \end{minipage}
\end{marginfigure}

The PlutoAR Android Application is used for viewing the augmented objects. The user points the camera of the mobile phone at the launchpad and starts placing tiles on the ``canvas'' area of the launchpad. For each tile that is placed on ``canvas'', the PlutoAR Interpreter recognises the action to be executed and the output appears in the launchpad. The immediate output generated is useful in debugging and understanding the logic step by step.  

	PlutoAR App is offline to protect the students from cyberbullying and to be accessible to children in places without an internet connectivity. 
Cyberbullying, addressed as ``An old problem in a new guise'' by Campbell \cite{campbell_2005}, is an important security factor to look into while students are given access to mobile devices. 
\\
According to ITU worldwide survey \cite{kawasumi_2004} on Rural Communications, more than 2.5 Billion people (40\% of world's population) live in the rural and remote areas of developing countries where access to telecommunications is still limited and with this in mind, we have designed it to work without internet connectivity.

We chose to make an Android application as Android has approximately 73.39\% of mobile operating system market share worldwide \cite{statista} and Android devices are cheaper as compared to other options.   
	As an example of a developing country, India has Android tablets priced as low as 35\$ \cite{raina_timmons_2011}. 

For implementation of the PlutoAR application, we used the Unity Game Engine \cite{unity} with C\# scripting, Android Software Development Kit (SDK) \cite{android}  and Vuforia AR SDK \cite{vuforia}. Vuforia facilitates easy and rapid implementation of marker based AR systems.
 
\section{Stories: AR Experiences}
PlutoAR has inbuilt AR experiences to stir up the intrinsic motivation of children and briefs them through, to explore and experiment with a variety of combinations with the toolkit.

Students can launch a rocket using the sequence of tiles, ``rocket'' and then ``takeoff''. A still image of this activity can be seen in Figure \ref{fig:marginfig1}. 

Similarly students can grow trees by placing the ``surface'' tile followed by ``tree'' and then growing them using ``rain'' \ref{fig:marginfig3}. Though these examples follow a simple algorithm to place the tiles, children can also experiment with any combination of the steps to learn and try out their own algorithms for example placing the ``surface'' and ``takeoff'' card could fly off the surface. This creates replay value and experimentation crucial in developing the motivation to use PlutoAR to play and learn.

Next, we look at an activity which focuses on computational paradigms such as looping and problem-solving. The ``Solving the Maze'' activity gives a simple goal to the child. Navigate the rocket to reach the planet through a maze of asteroids with movement and loop tiles. The child could move the rocket forward by one step or use a loop tile to move it five steps forward. This will differ from child to child. For example,  some children may think of  different path to navigate the rocket. This activity brings out the computational thinking and encourages children to develop the intuition of algorithmic design.

Finally, we look at the ``Asteroid Math'' activity where children have to add or subtract asteroids to solve a math equation as visible in Figure \ref{fig:marginfig4}. PlutoAR  checks if the answer is correct. This teaches basic elementary mathematics through playing and interactive story telling. 

All these AR Experiences are demonstrated in the demo video. \cite{demovideo}

\section{School Visits}

We visited KIIT International School, Odisha, India, to test our product in the real world scenario and the feedbacks, we received, from the children were encouraging . The children actively participated in the demonstration activity which went to show how effective AR in education could be.
 \begin{figure}[H]
   \includegraphics[width=1\columnwidth]{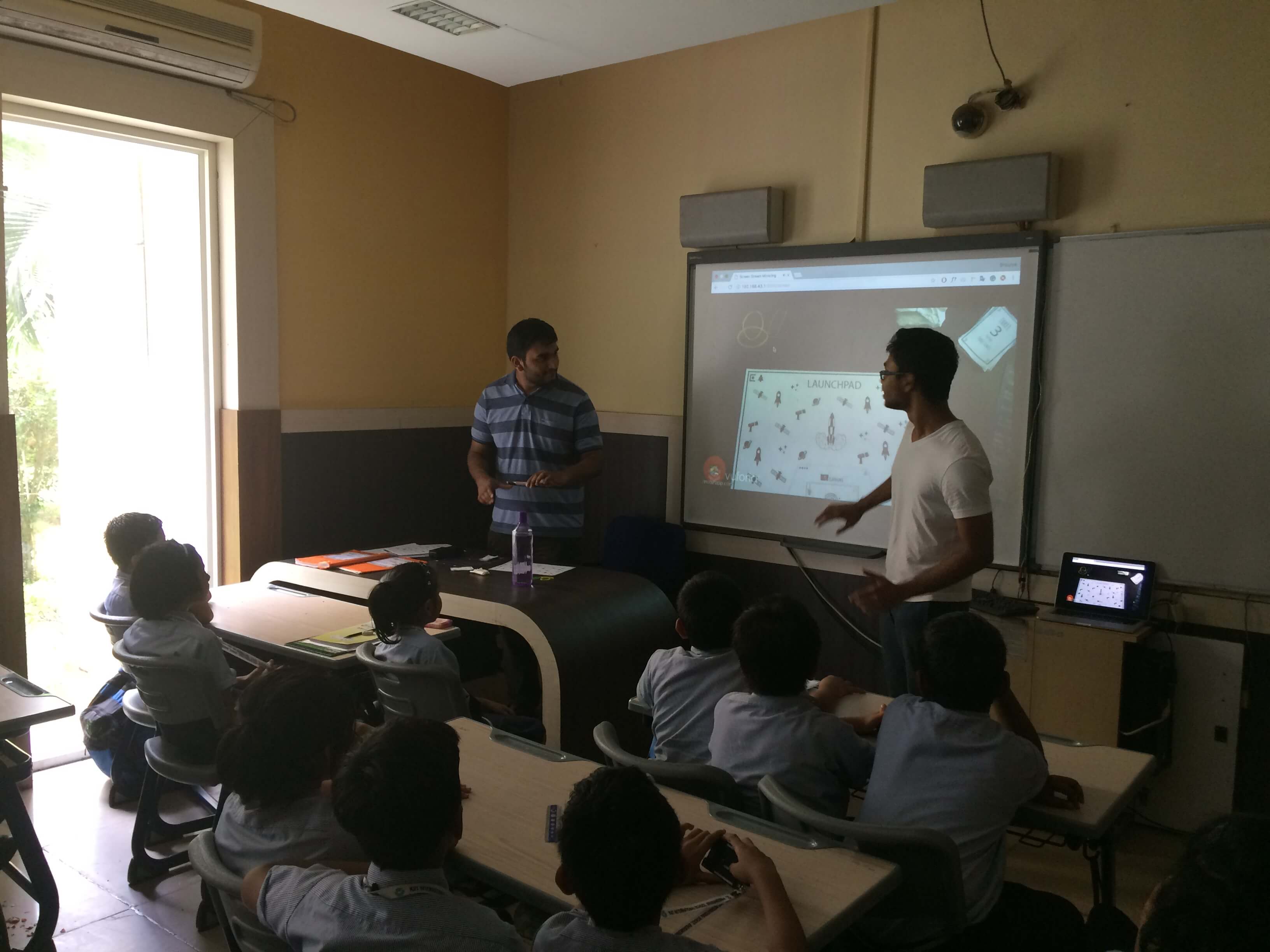}
   \caption{Demonstration of PlutoAR application at KIIT International School, Bhubaneswar, Odisha, India.}\label{fig:bats}
 \end{figure}
\section{Conclusion and Future work}	
PlutoAR is an inexpensive, tangible and portable educational toolkit which manages to move forward a step to teach children to program, play and learn educational facts.
The AR experiences are meticulously designed to introduce computational thinking in children before core computer courses. Regarding STEM, PlutoAR's fact system provides small yet summarized and important facts about science and mathematics.
Above all, the purpose of PlutoAR is to make children intrinsically motivated to learn programming through play.
\par
The current version of PlutoAR is at a prototypical stage. In the next version, we plan to include the school curriculum in the form of AR experiences in PlutoAR. We may also experiment different interaction possibilities with the virtual objects, such as gesture and gaze-based interactions and look at methods to expand this to kids who require special attention. 

\bibliographystyle{SIGCHI-Reference-Format}
\bibliography{PlutoAR}

\end{document}